\documentclass{amsart}

 \usepackage[foot]{amsaddr}
\usepackage{amsmath, bbm}
\usepackage{amsthm}
\usepackage{mathabx}
\usepackage{fullpage}

\usepackage{asymptote}
\usepackage{enumerate} 
\usepackage{setspace} 
\usepackage{subcaption} 
\usepackage[
pdftitle={Quantum LCP Solver},
pdfcreator={pdflatex},
pdfsubject={preprint},
hyperindex = {true},
colorlinks = {true},
linkcolor = {blue},
citecolor = {blue}
]{hyperref}
\usepackage{array}
\usepackage{cancel}
\usepackage{stmaryrd}
\usepackage{booktabs}
\usepackage[dvips]{epsfig}
\usepackage{epstopdf}
\usepackage{setspace}
\usepackage{lipsum}
\usepackage[shortlabels]{enumitem}
\usepackage{algorithm, algpseudocode}

\usepackage{cleveref}
\crefformat{equation}{(#2#1#3)}

\newcommand{\abs}[1]{\left\lvert #1 \right\rvert}
\newcommand{\norm}[1]{\left\lVert #1 \right\rVert}
\newcommand{\ip}[1]{\left\langle #1 \right\rangle}
\newcommand{\paren}[1]{\left( #1 \right)}
\newcommand{\braces}[1]{\left\{ #1 \right\}}
\newcommand{\bracket}[1]{\left[ #1 \right]}

\newcommand{\ket}[1]{\left\lvert #1 \right\rangle}
\newcommand{\ketbra}[2]{\mathinner{|{#1}\rangle\langle{#2}|}}

\newcommand{\D}{\mathrm{d}}

\usepackage[normalem]{ulem}
\usepackage[dvipsnames]{xcolor}
\usepackage{etoolbox}

\title{Variational quantum and neural quantum states algorithms for the linear complementarity problem}

\begin{document}
% \author[1]{Saibal De}
% \author[2,*]{Oliver Knitter}
% \author[2]{Rohan Kodati}
% \author[3]{Paramsothy Jayakumar}
% \author[2]{ \\ James Stokes}
% \author[2]{Shravan Veerapaneni}

% \affil[1]{Sandia National Laboratories}
% \affil[2]{Department of Mathematics, University of Michigan}
% \affil[3]{US Army GVSC}
% \affil[*]{Corresponding author; knitter@umich.edu}
%
%
\author{Saibal De$^{1}$}
\author{Oliver Knitter$^{2 *}$}
\author{Rohan Kodati$^{2}$}
\author{Paramsothy Jayakumar$^{3}$}
\author{\vspace{0.05in} \\ James Stokes$^{2}$}
\author{Shravan Veerapaneni$^{2}$}
\address{$^1$Sandia National Laboratories, Livermore, CA 94551, USA}
\address{$^2$Department of Mathematics, University of Michigan, Ann Arbor, MI 48109, USA}
\address{$^3$Ground Vehicle Systems Center, U.S. Army DEVCOM, Warren, MI}
\thanks{$^*$Corresponding author. \emph{Email address:} knitter@umich.edu \vspace{0.05in} \\ DISTRIBUTION STATEMENT A. Approved for public release; distribution is unlimited. \\
OPSEC$\# 9098$.
%distribution unlimited. OPSEC $\#$
}
\date{}

\begin{abstract}
    Variational quantum algorithms (VQAs) are promising hybrid quantum-classical methods designed to leverage the computational advantages of quantum computing while mitigating the limitations of current noisy intermediate-scale quantum (NISQ) hardware. Although VQAs have been demonstrated as proofs of concept, their practical utility in solving real-world problems---and whether quantum-inspired classical algorithms can match their performance---remains an open question. We present a novel application of the variational quantum linear solver (VQLS) and its classical neural quantum states-based counterpart, the variational neural linear solver (VNLS), as key components within a minimum map Newton solver for a complementarity-based rigid body contact model. We demonstrate using the VNLS that our solver accurately simulates the dynamics of rigid spherical bodies during collision events. These results suggest that quantum and quantum-inspired linear algebra algorithms can serve as viable alternatives to standard linear algebra solvers for modeling certain physical systems.
\end{abstract}

\maketitle

\noindent
{\small Keywords: Variational Quantum Algorithms, Neural Quantum States, Physical Simulation}

\section{Introduction}
\label{sec:intro}

High-fidelity physics-based models are routinely employed to study the mobility of ground vehicles on both on-road and off-road terrains \cite{marple2021active, mechergui2020efficient, wasfy2019understanding}. In these simulations, the terrain is represented as a granular medium composed of a large number of soil particles. The discrete element method (DEM) tracks the individual physical states---namely, position, velocity, and force---of each particle by solving a system of ordinary differential equations (ODEs) derived from Newton's laws of motion. Macroscopic soil properties such as density and elasticity emerge from the collective responses of these particles to body forces such as gravity and contact interactions. While these simulations generate accurate predictions of vehicle-terrain systems, they are computationally expensive due to the large number of particles involved.

A key component of DEM simulations is the calculation of pairwise contact forces between particles. Penalty-based DEM simulations (DEM-P) introduce contact force fields that repel particles when they come into close proximity, preventing interpenetration \cite{berger2015hybrid, mazhar2013chrono}. These penalty forces are simple to implement and computationally inexpensive to evaluate. However, they introduce artificial stiffness into the system of ODEs in the form of spring-like forces. Accurately modeling rigid particle interactions requires very high spring coefficients to represent sharp contacts, which significantly limits the viable time step sizes for numerical ODE solvers. To address this limitation, complementarity-based DEM simulations (DEM-C) instead consider the geometric constraints of the contact problem \cite{anitescu1997formulating, anitescu2010iterative, stewart1996implicit}. In this approach, the contact forces are shown to satisfy complementarity conditions corresponding to certain optimization problems, allowing the forces to be obtained by solving convex programs.

First-order optimizers are attractive for solving smooth convex programs because they only require gradient evaluations which, in the context of collisions,  reduce to sparse matrix-vector products. Unfortunately, convergence rates of these solvers can often be quite slow. In contrast, second-order solvers, which incorporate the Hessian matrix of the objective function, offer faster convergence at the expense of repeatedly solving linear systems \cite{de2019scalable}. For non-smooth objective functions, such as those arising in minimum map methods \cite{niebe2022numerical} for DEM-C convex programs, optimizers employing Newton's root finding algorithm also need to solve specific linear systems.

The computational cost of solving these large, sparse linear systems is the main bottleneck in applying robust optimization algorithms to collision force calculations in DEM-C models of granular media. Quantum and quantum-inspired machine learning paradigms offer a potential solution to this scalability barrier. For instance, the variational quantum linear solver (VQLS) \cite{bravo2019variational}, an adaptation of the variational quantum eigensolver (VQE), is a variational quantum algorithm (VQA) capable of solving sufficiently sparse linear systems of exponentially large dimension. As a near-term algorithm, the VQLS can run on existing noisy intermediate-scale quantum (NISQ) hardware; unlike more theoretically promising quantum linear system solvers \cite{harrow2009quantum}, it does not require further quantum hardware development to be viable. Hypothetically, it could serve as a black-box linear solver for collision forces within DEM-C, demonstrating a worthwhile practical application of VQAs.

At the same time, VQAs present their own challenges. Unlike fault-tolerant algorithms, they are not known to possess quantum advantage. Indeed, neural quantum states (NQS) \cite{carleo2017solving}---a variational Monte Carlo deep learning algorithm sharing close parallels with the VQE---has been shown to provide a viable pathway for ``de-quantizing'' certain VQAs \cite{knitter2022vnls}. NQS has already demonstrated significant success in solving some problems addressed by VQAs, particularly \textit{ab initio} quantum chemistry~\cite{Choo_2020}, combinatorial optimization \cite{gomes2019classical} and linear systems via the variational neural linear solver (VNLS) \cite{knitter2022vnls}. Thus the VNLS may in its own right comprise a scalable black box solver within the DEM-C framework. Additionally, the structure of VQAs is inherently defined by the limitations placed on quantum hardware, particularly through the restrictive ways in which information can be encoded into and retrieved from a quantum circuit. NQS, a classical method that only parallels VQAs, can pose greater flexibility in this regard, making the VNLS and similar NQS-based dequantizations more practical for certain tasks. Previous work on VQLS and VNLS has been limited to Ising model--inspired toy problems \cite{bravo2019variational, knitter2022vnls}, so while these methods are promising, the qualitative differences in performance between them remain unclear, especially when considering practical applications.

In the hopes of encouraging further research into physical simulation as an area of application for quantum and quantum-inspired machine learning methods, this paper explores the potential utility of the VNLS and, to a more hypothetical extent, the VQLS, as black box solvers for obtaining DEM-C collision forces. We start by directly comparing the VQLS and the VNLS, applied under analogous training conditions, to Ising-inspired baseline linear systems. Next we apply VNLS to a non-Ising system, derived from a rudimentary granular medium simulation, and present preliminary results on applicability of VQLS to DEM-C simulators.

The rest of this article is structured as follows. In Section \ref{sec:lcp}, we review the DEM-C formulation for rigid body contact and derive the specific linear systems that need to be solved within our framework. In Section \ref{sec:vqa}, we provide an overview of the VQLS and VNLS, and describe the general framework for incorporating these models into a DEM-C simulator. In Section \ref{sec:result}, we compare the two solvers on baseline problems, and present preliminary evidence of the VNLS's suitability as a DEM-C solver using proof-of-concept granular media simulations. Specifically, we simulate a small number of rigid particles in free fall inside a hollow, rigid spherical boundary, neglecting for now the influence of frictional forces, and demonstrate that the VNLS exhibits suitable performance when solving the complementarity problems that arise. We also briefly explore the primary bottleneck in using VQLS within the DEM-C framework, namely that the Pauli-string decomposition of the linear systems produces a large number of terms. Section \ref{sec:conclusion} summarizes the content of this paper and discusses potential avenues for further exploration.

\section{Primer on DEM-C Formulation of Rigid Body Contact}
\label{sec:lcp}

In this section, we briefly review the complementarity formulation of rigid body contact following \cite{stewart1996implicit, anitescu1997formulating}. To help make this exposition more succinct, we only consider frictional contact between rigid spheres, ignoring any rotational degrees of freedom. We derive the actual linear complementarity problem (LCP) by linearizing the Coulomb friction model, and introduce the Newton minimum-map solver. We end the section by explicitly writing out the sparse linear system that we need to solve to obtain the contact forces.

\subsection{Equations of Motion}

We consider the motion of a system of $n_b$ spherical rigid bodies of uniform density in three-dimensional space. Let $p_i = (p_{i,1}, p_{i,2}, p_{i,3}) \in \mathbb{R}^3$ and $v_i = (v_{i,1}, v_{i,2}, v_{i,3}) \in \mathbb{R}^3$ denote the position and velocity of the center of mass of the $i$-th sphere, and let $m_i$ be its mass. The generalized position $p = (p_1, \ldots, p_{n_b}) \in \mathbb{R}^{3 n_b}$ and generalized velocity $v = (v_1, \ldots, v_{n_b}) \in \mathbb{R}^{3 n_b}$ of the global system evolves according to Newton's laws of motion:
\begin{equation}
    \label{eq:ode_global}
    \frac{\D p}{\D t} = v, \quad M \frac{\D v}{\D t} = f + f^\text{col}.
\end{equation}
Here $f$ and $f^\text{col}$ are the net external force---in this case, gravity---and the net inter-particle collision force; $M = \operatorname{diag}(m_1, m_1, m_1, \ldots, m_{n_b}, m_{n_b}, m_{n_b}) \in \mathbb{R}^{3 n_b \times 3 n_b}$ is the generalized mass matrix.

\begin{figure}
    \centering
    \begin{subfigure}{0.49\textwidth}
        \centering
        \includegraphics[scale=0.35]{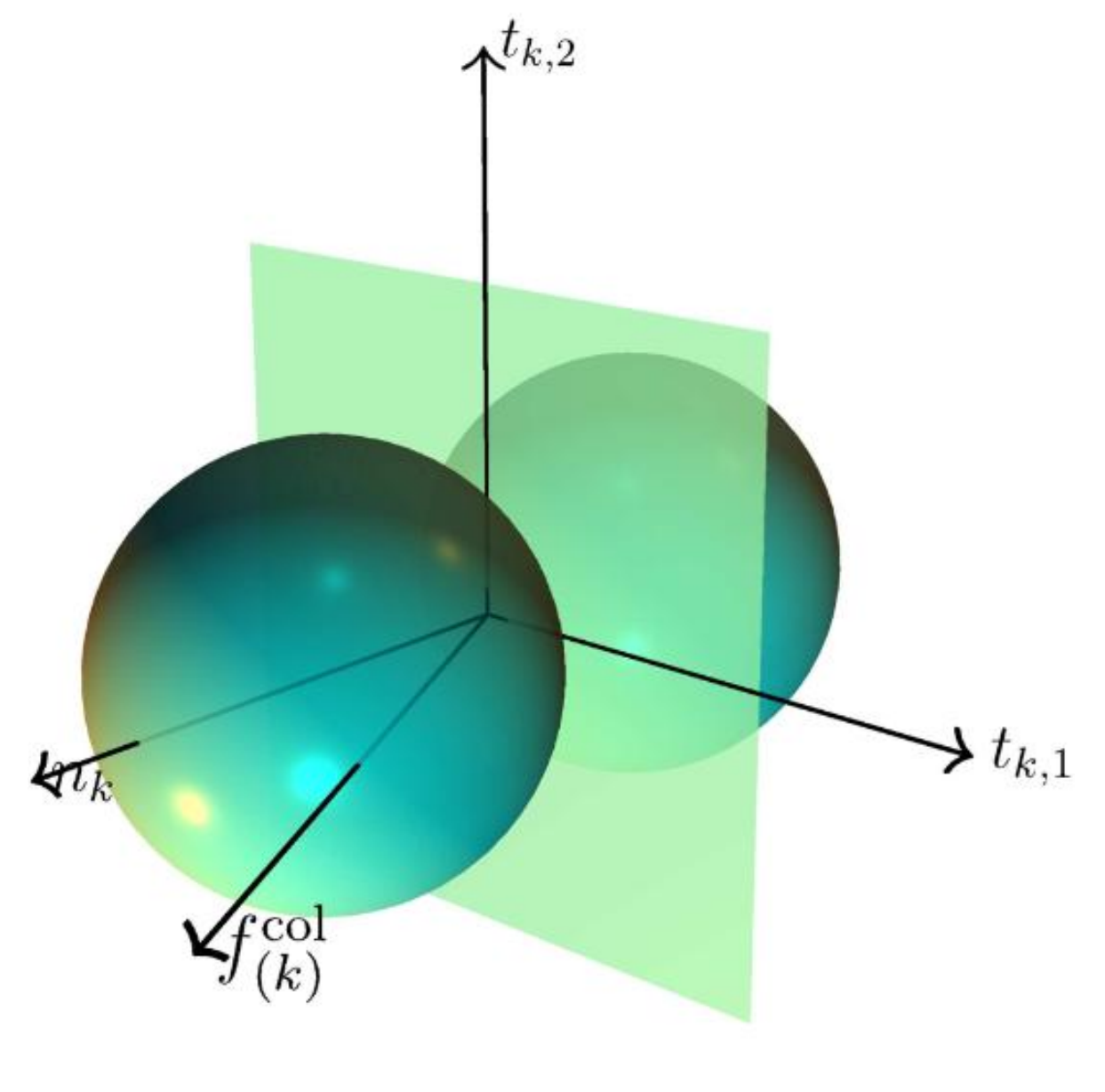}
        \caption{\em The local coordinate frame for the $k$-th contact between two spheres. The vector $\nu_k$ is the unit normal to the contact plane, while $\tau_{k,1}$ and $\tau_{k,2}$ form an orthonormal basis for the contact plane. The contact force is decomposed along these three directions as $f_{(k)}^\text{col} = \gamma_{k,n} \nu_k + \gamma_{k,1} \tau_{k,1} + \gamma_{k,2} \tau_{k,2}$.}
        \label{fig:sphere_collision}
    \end{subfigure}
    \hfill
    \begin{subfigure}{0.44\textwidth}
        \centering
        \includegraphics[scale=0.33]{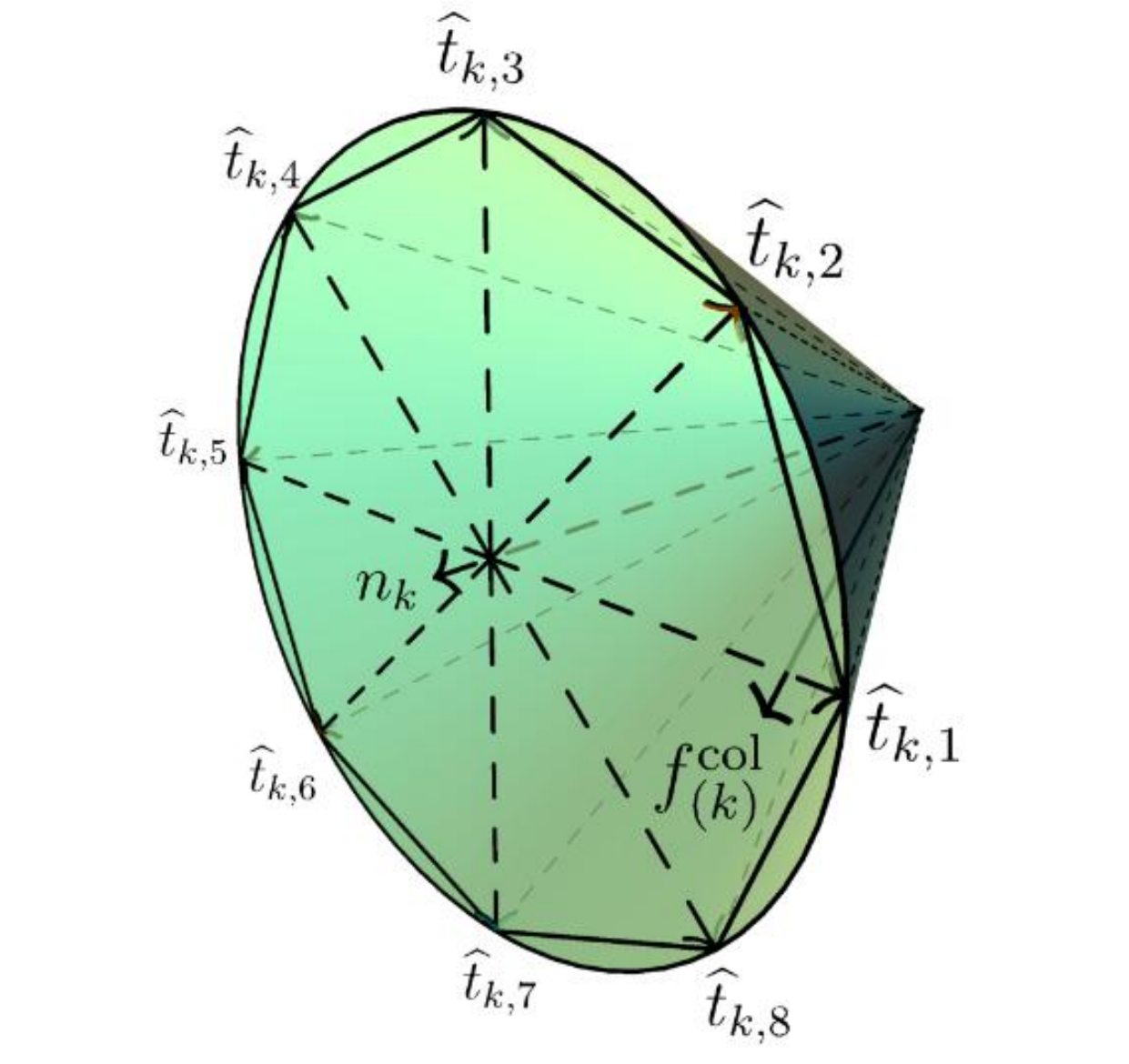}
        \caption{\em The collision force $f_{(k)}^\text{col}$ of contact $k$ lies within the Coulomb friction cone. We linearly approximate this quadratic Coulomb cone using a polyhedral cone with octagonal base. The direction vectors $\{\widehat{\tau}_{k,\ell} : 1 \leq \ell \leq 8\}$ are balanced and positively span the contact plane.}
        \label{fig:cone_linearization}
    \end{subfigure}
    \caption{\em The local coordinate frame for the collision between two spheres, and the linearization of the corresponding Coulomb friction cone.}
\end{figure}

We construct an expression for the collision force as follows. Let $n_c$ denote the number of pairwise contacts between the spheres, and suppose the $k$-th contact occurs between spheres $i_k$ and $j_k$ (we assume $i_k < j_k$ without loss of generality). Then $\nu_k = (x_{j_k} - x_{i_k}) / \norm{x_{j_k} - x_{i_k}}$ is the unit vector normal to the contact plane and pointed toward $j_k$-th sphere. Suppose $\tau_{k,1}, \tau_{k,2} \in \mathbb{R}^3$ form an orthonormal basis for the contact plane (see \cref{fig:sphere_collision}), and decompose the contact force $f^\text{col}_{(k)}$ active on the $j_k$-th sphere from the $k$-th collision along these orthonormal directions as
\begin{equation*}
    f^\text{col}_{(k)} = \gamma_{k, \nu} \nu_k + \gamma_{k, 1} \tau_{k, 1} + \gamma_{k, 2} \tau_{k, 2}.
\end{equation*}
Denote by $d_{k,\nu}, d_{k,1}, d_{k,2} \in \mathbb{R}^{3 n_b}$ the generalized direction vectors corresponding to the three axes $\nu_k, \tau_{k,1}, \tau_{k,2}$ of the local coordinate system; see \cref{app:fcol_global} for the precise construction. Then the total contact force is given by
\begin{equation}
    \label{eq:fcol_global}
    f^\text{col} = \sum_{k = 1}^{n_c} (\gamma_{k,\nu} d_{k,\nu} + \gamma_{k,1} d_{k,1} + \gamma_{k,2} d_{k,2}) = D \gamma,
\end{equation}
where we define
\begin{align*}
    D =
    \begin{bmatrix}
        \cdots & d_{k,\nu} & d_{k,1} & d_{k,2} & \cdots
    \end{bmatrix}
    \in \mathbb{R}^{3 n_b \times 3 n_c},
    \quad
    \gamma = (\ldots, \gamma_{k,\nu}, \gamma_{k,1}, \gamma_{k,2}, \ldots) \in \mathbb{R}^{3 n_c}.
\end{align*}

We can evolve the system of ODEs described in \cref{eq:ode_global} forward in time using standard numerical integrators, provided we have a model for computing the contact force magnitudes $\gamma$. In the subsequent sections, we use the DEM-C framework to reformulate $\gamma$ as a solution to a linear complementarity problem.

\subsection{Coulomb Friction Model}

The Coulomb model establishes a relationship between the magnitude of the friction force (i.e., tangential part of the contact force) $f_{(k)}^\text{fric} = \gamma_{k,1} \tau_{k,1} + \gamma_{k,2} \tau_{k,2}$, the magnitude of the corresponding normal contact force $\gamma_{k,\nu}$, and the velocity of the system. Specifically, the components of friction force satisfy the maximal dissipation condition
\begin{equation}
    \label{eq:coulomb_dissipation}
    (\gamma_{k,1}, \gamma_{k,2}) = \operatorname*{argmin}_{\sqrt{\widehat{\gamma}_{k,1}^2 + \widehat{\gamma}_{k,2}^2} \leq \mu \gamma_{k,\nu}} (\widehat{\gamma}_{k,1} d_{k,1} + \widehat{\gamma}_{k,2} d_{k,2})^\top v,
\end{equation}
where $\mu$ is the coefficient of friction between the spheres. See \cref{app:coulomb} for a derivation of this condition.

We interpret the constraint in \cref{eq:coulomb_dissipation} as the contact force components vector $(\gamma_{k,\nu}, \gamma_{k,1}, \gamma_{k,2})$ lying within a convex cone with half-angle $\arctan(\mu)$, referred to as the Coulomb friction cone. In practice, this convex cone is sometimes approximated by a polyhedral cone in order to keep the problem linear. This linearization corresponds to an approximation of the friction force,
\begin{equation*}
    f_{(k)}^\text{fric} \approx \beta_{k,1} \widehat{\tau}_{k,1} + \cdots + \beta_{k,k} \widehat{\tau}_{k,s},
\end{equation*}
where the coefficients $\beta_{k,\ell}$ are constrained to be nonnegative. As depicted in \cref{fig:cone_linearization}, the direction vectors $\widehat{\tau}_{k,1}, \ldots, \widehat{\tau}_{k,s}$ positively span the contact plane and are balanced: for any $1 \leq \ell \leq s$, there exists $1 \leq \ell' \leq s$ such that $\widehat{\tau}_{k,\ell'} = -\widehat{\tau}_{k,\ell}$. Let $\widehat{d}_{k,1}, \ldots, \widehat{d}_{k,s}$ be the generalized global direction vectors associated with local direction vectors $\widehat{\tau}_{k,1}, \ldots, \widehat{\tau}_{k,s}$; see \cref{app:fcol_global} for the exact construction. In place of \cref{eq:fcol_global}, the contact force in the ODE system \cref{eq:ode_global} is now modeled as
\begin{equation*}
    f^\text{col} = \sum_{k = 1}^{n_c} (\gamma_{k,\nu} d_{k,\nu} + \beta_{k,1} \widehat{d}_{k,1} + \cdots + \beta_{k,s} \widehat{d}_{k,s}) = D_\nu \gamma_\nu + \widehat{D} \beta.
\end{equation*}
where we separate out the normal and tangential contact entities:
\begin{align*}
    D_\nu &=
    \begin{bmatrix}
        d_{1,\nu} & \cdots & d_{n_c,\nu}
    \end{bmatrix}
    \in \mathbb{R}^{3 n_b \times n_c},
    &
    \gamma_\nu &= (\gamma_{1,\nu}, \ldots, \gamma_{n_c,\nu})
    \in \mathbb{R}^{n_c},
    \\
    \widehat{D}_k &=
    \begin{bmatrix}
        \widehat{d}_{k,1} & \cdots & \widehat{d}_{k,s}
    \end{bmatrix}
    \in \mathbb{R}^{3 n_b \times s},
    &
    \beta_k &= (\beta_{k,1}, \ldots, \beta_{k,s})
    \in \mathbb{R}^s,
    \\
    \widehat{D} &=
    \begin{bmatrix}
        \widehat{D}_1 & \cdots & \widehat{D}_{n_c}
    \end{bmatrix}
    \in \mathbb{R}^{3 n_b \times s n_c},
    &
    \beta &= (\beta_1, \ldots, \beta_{n_c})
    \in \mathbb{R}^{s n_c};
\end{align*}
In this linearized approximation, the Coulomb dissipation condition \cref{eq:coulomb_dissipation} reduces to
\begin{equation}
    \label{eq:linearized_coulomb_dissipation}
    \beta_k = \operatorname*{argmin}_{\substack{\widehat{\beta}_k \geq 0 \\ e^\top \widehat{\beta}_k \leq \mu \gamma_{k,\nu}}} (\widehat{\beta}_{k,1} \widehat{d}_{k,1} + \cdots + \widehat{\beta}_{k,s} \widehat{d}_{k,s})^\top v, \quad e = (1, \ldots, 1) \in \mathbb{R}^s.
\end{equation}

\subsection{Linear Complementarity Formulation}
\label{sec:lcp_formulation}

Note that when the $k$-th pair of particles are in contact, the magnitude of the normal contact force may be positive, but the relative velocity between the particles in the normal direction must be zero. Conversely, once the collision ends, the contact force is zero but the relative velocity along the normal direction must be non-negative. Given that the normal component of relative velocity between the $k$-th pair of spheres equals $d_{k,\nu}^\top v$ (see \cref{app:coulomb}), we write
\begin{equation*}
    \gamma_{k,\nu} \geq 0, \quad d_{k,\nu}^\top v \geq 0, \quad \gamma_{k,\nu} \cdot (d_{k,\nu}^\top v) = 0.
\end{equation*}
We symbolically express this set of conditions as the complementarity condition
\begin{equation*}
    0 \leq \gamma_{k,\nu} \perp d_{k,\nu}^\top v \geq 0.
\end{equation*}
The Karush-Kuhn-Tucker (KKT) conditions for the maximal dissipation law of the linearized Coulomb friction model \cref{eq:linearized_coulomb_dissipation} are similarly expressed as complementarity conditions
\begin{align*}
    0 \leq \beta_k &\perp \lambda_k e + \widehat{D}_k^\top v \geq 0, \\
    0 \leq \lambda_k &\perp \mu \gamma_{k,\nu} - e^\top \beta_k \geq 0,
\end{align*}
where $\lambda_k$ denotes slack variables introduced for the inequality constraints. These three sets of linear complementarity equations define the contact force magnitudes $\gamma_{k,\nu}$ and $\beta_k$. In the next section, we will combine them to form a single LCP in terms of the time-discretized particle states.

\subsection{Time Discretization}

\begin{algorithm}[t]
    \caption{Time evolution of a system of rigid bodies within DEM-C framework}
    \label{alg:euler_update}
    \begin{algorithmic}[1]
        \State {\textbf{global} Mass matrix $M$, coefficient of friction $\mu$}
        \Function{EulerUpdate}{current position $p^{(\ell)}$, current velocity $v^{(\ell)}$, timestep size $h$}
            \State {Evaluate external force $f^{(\ell)}$ from current position and velocity of the rigid body system}
            \State {Evolve velocity ignoring collision: $v^{(\ell + 1)}_\text{known} \gets v^{(\ell)} + h M^{-1} f^{(\ell)}$}
            \State {Detect particles that are expected to collide from current particle positions}
            \State {Compute direction vectors for the contacts and assemble the $D_\nu^{(\ell)}$ and $\widehat{D}^{(\ell)}$ matrices}
            \State {Assemble the data matrix $Q$ of the LCP as defined in \cref{eq:lcp}}
            \State {Assemble the RHS vector $r$ of the LCP as defined in \cref{eq:lcp}}
            \State {Obtain solution: $y \gets \Call{SolveLCP}{Q, r, 0}$} \Comment{See \cref{alg:lcp_solve}}
            \State {Read off normal and frictional contact impulse magnitudes, $\gamma_\nu^{(\ell + 1)}$ and $\beta^{(\ell + 1)}$, from solution $y$}
            \State {Update velocity: $v^{(\ell + 1)} \gets v^{(\ell + 1)}_\text{known} + M^{-1} (D_\nu^{(\ell)} \gamma_\nu^{(\ell + 1)} + \widehat{D}^{(\ell)} \beta^{(\ell + 1)})$}
            \State {Update position: $p^{(\ell + 1)} \gets p^{(\ell)} + h v^{(\ell + 1)}$}
            \State {\Return Updated position $p^{(\ell + 1)}$ and velocity $v^{(\ell + 1)}$}
        \EndFunction
    \end{algorithmic}
\end{algorithm}

We discretize the complementarity conditions using a mixed forward--backward Euler scheme; using superscripts to denote timestep indices, we write complementarity conditions for the $\ell$-th timestep as
\begin{align*}
    0 \leq \gamma_{k,\nu}^{(\ell + 1)} &\perp d_{k,\nu}^{(\ell),\top} v^{(\ell + 1)} \geq 0, \\
    0 \leq \beta_k^{(\ell + 1)} &\perp \lambda_k^{(\ell + 1)} e + \widehat{D}_k^{(\ell), \top} v^{(\ell + 1)} \geq 0, \\
    0 \leq \lambda_k^{(\ell + 1)} &\perp \mu \gamma_{k,\nu}^{(\ell + 1)}  - e^\top \beta_k^{(\ell + 1)} \geq 0,
\end{align*}
where the velocity vector $v^{(\ell + 1)}$ corresponding to the next timestep is given by
\begin{align}
    \label{eq:velocity_update}
    v^{(\ell + 1)} &= \underbrace{v^{(\ell)} + h M^{-1} f^{(\ell)}}_{v_\text{known}^{(\ell + 1)}} + M^{-1} (D_\nu^{(\ell)} \gamma_\nu^{(\ell + 1)} + \widehat{D}^{(\ell)} \beta^{(\ell + 1)}).
\end{align}
Note that we have absorbed the timestep size $h$ into the $\beta$ and $\gamma_\nu$ terms---turning them from contact forces to contact impulses---for better numerical stability. We can interpret this velocity update in \cref{eq:velocity_update} as follows: first compute the ``known'' velocity of the particles by using a forward Euler step with only the external forces, then correct it to adjust for any collisions. Combining these equations, we obtain a single LCP
\begin{equation}
    \label{eq:lcp}
    0 \leq
    \underbrace{
    \begin{bmatrix}
        \gamma_\nu^{(\ell + 1)} \\
        \beta^{(\ell + 1)} \\
        \lambda^{(\ell + 1)}
    \end{bmatrix}
    }_{y \in \mathbb{R}^{(s + 2) n_c}}
    \perp
    \underbrace{
    \begin{bmatrix}
        D_\nu^{(\ell),\top} M^{-1} D_\nu^{(\ell)} & D_\nu^{(\ell),\top} M^{-1} \widehat{D}^{(\ell)} & 0 \\
        \widehat{D}^{(\ell),\top} M^{-1} D_\nu^{(\ell)} & \widehat{D}^{(\ell),\top} M^{-1} \widehat{D}^{(\ell)} & E \\
        \mu I_{n_c} & -E^\top & 0
    \end{bmatrix}
    }_{Q \in \mathbb{R}^{(s + 2) n_c \times (s + 2) n_c}}
    \underbrace{
    \begin{bmatrix}
        \gamma_\nu^{(\ell + 1)} \\
        \beta^{(\ell + 1)} \\
        \lambda^{(\ell + 1)}
    \end{bmatrix}
    }_{y \in \mathbb{R}^{(s + 2) n_c}}
    +
    \underbrace{
    \begin{bmatrix}
        D_\nu^{(\ell),\top} v^{(\ell + 1)}_\text{known} \\
        \widehat{D}^{(\ell),\top} v^{(\ell + 1)}_\text{known} \\
        0
    \end{bmatrix}
    }_{r \in \mathbb{R}^{(s + 2) n_c}}
    \geq 0,
\end{equation}
where $I_{n_c}$ is the $n_c \times n_c$ identity matrix, and $E = \operatorname{diag}(e, \ldots, e) \in \mathbb{R}^{s n_c \times n_c}$. Let $m = (s + 2) n_c$ denote the size of the LCP, with $Q \in \mathbb{R}^{m \times m}$ and $r \in \mathbb{R}^m$.

Once we solve for $y$ in the above LCP, we obtain the contact impulse magnitudes $\gamma_\nu$ and $\beta$ for timestep $\ell + 1$. We use these values to first update the velocity by applying \cref{eq:velocity_update}, then finally update the position as
\begin{equation*}
    p^{(\ell + 1)} = p^{(\ell)} + h v^{(\ell + 1)}.
\end{equation*}
In \cref{alg:euler_update}, we list the key steps of the full Euler scheme for the rigid body system.

\subsection{Minimum Map Newton Solver}

To complete our description of the Euler scheme, we must still give a solution for the linear complementarity problem \cref{eq:lcp}. Our approach is based on a minimum map Newton method for LCPs \cite{niebe2022numerical}. For $Q \in \mathbb{R}^{m \times m}$ and $r \in \mathbb{R}^m$, define the minimum map $\phi : \mathbb{R}^m \to \mathbb{R}^m$ by
\begin{equation}
    \label{eqn:min_map}
    \phi(y) =
    (\phi_1(y_1, z_1), \ldots, \phi_m(y_m, z_m)),
    \quad z = Q y + r, \quad \phi_i(y_i, z_i) = \min\{y_i, z_i\}.
\end{equation}
We observe that any root of \cref{eqn:min_map} provides a solution to the LCP.
Using Newton's method, we start at a specified value of $y^{(\ell)}$ and update it by $y^{(\ell + 1)} \gets y^{(\ell)} + \Delta y$, where the increment $\Delta y$ satisfies
\begin{equation}
    \label{eq:newton}
    0 = \phi(y^{(\ell + 1)}) \approx \phi(y^{(\ell)}) + \Phi(y^{(\ell)}) \Delta y, \quad \Phi_{ij}(y^{(\ell)}) = \frac{\partial \phi_i}{\partial y_j}(y^{(\ell)}).
\end{equation}
Note that each component function $\phi_i$ is non-smooth, so the partial derivatives $\frac{\partial \phi_i}{\partial y_j}$ are taken in a generalized sense; a simple choice is constructed by observing that
\begin{equation*}
    \phi_i(y_i) =
    \begin{cases}
        z_i = (Q y + r)_i & \text{if} \quad z_i < y_i, \\
        y_i & \text{if} \quad z_i \geq y_i,
    \end{cases}
    \implies
    \frac{\partial \phi_i}{\partial y_j} =
    \begin{cases}
        q_{ij} & \text{if} \quad z_i < y_i, \\
        1 & \text{if} \quad z_i \geq y_i \text{ and } j = i, \\
        0 & \text{if} \quad z_i \geq y_i \text{ and } j \neq i,
    \end{cases}
\end{equation*}
where $q_{ij}$ is the $(i, j)$-th entry of the matrix $Q$. With this structure of the Hessian matrix $\Phi(y^{(\ell)})$, we define the index partition
\begin{equation*}
    \mathcal{A} = \{i : 1 \leq i \leq m, \; z_i^{(\ell)} < y_i^{(\ell)}\}, \quad \mathcal{B} = \{i : 1 \leq i \leq m, \; z_i^{(\ell)} \geq y_i^{(\ell)}\};
\end{equation*}
after applying a suitable permutation, we rewrite \cref{eq:newton} as
\begin{equation}
    \label{eq:lcp_linear_system}
    0 =
    \begin{bmatrix}
        \phi_\mathcal{A}(y^{(\ell)}) \\
        \phi_\mathcal{B}(y^{(\ell)})
    \end{bmatrix}
    +
    \begin{bmatrix}
        Q_{\mathcal{A}\mathcal{A}} & Q_{\mathcal{A} \mathcal{B}} \\
        0 & I_{|\mathcal{B}|}
    \end{bmatrix}
    \begin{bmatrix}
        \Delta y_{\mathcal{A}} \\
        \Delta y_{\mathcal{B}}
    \end{bmatrix}
    \implies
    Q_{\mathcal{A} \mathcal{A}} \Delta y_{\mathcal{A}} = Q_{\mathcal{A}\mathcal{B}} \phi_\mathcal{B}(y^{(\ell)}) - \phi_\mathcal{A}(y^{(\ell)}),
\end{equation}
where the index set subscript implies the construction of a sub-vector or sub-matrix with those specific entries from $\Delta y$, $\phi$ or $Q$, and $I_{|\mathcal{B}|}$ is the identity matrix of size equaling the number of indices in $\mathcal{B}$.

In \cref{alg:lcp_solve}, we outline the steps of the minimum-map Newton optimizer for the LCP. We note that the solution of the linear system in \cref{eq:lcp_linear_system} is the most expensive step of the algorithm; therefore it is also the main bottleneck in the DEM-C simulation of granular media. For this reason, we consider both the quantum VQLS and the quantum-inspired VNLS as potential black box solvers for \cref{eq:lcp_linear_system}. This work explores the possibility that these algorithms could relieve the computational bottleneck of, and thereby adding increased stability to, the DEM-C simulation framework.

\begin{algorithm}[t]
    \caption{Minimum-map Netwon solver for LCP}
    \label{alg:lcp_solve}
    \begin{algorithmic}[1]
        \Function{SolveLCP}{Matrix $Q \in \mathbb{R}^{m \times m}$, RHS $b \in \mathbb{R}^m$, initial guess $y^{(0)} \in \mathbb{R}^m$}
            \State {Initialize $\ell \gets 0$}
            \Repeat
                \State {Compute $z^{(\ell)} \gets A y^{(\ell)} + b$}
                \State {Construct the index sets $\mathcal{A} \gets \{i : 1 \leq i \leq m, \; z_i^{(\ell)} < y_i^{(\ell)}\}$ and $\mathcal{B} \gets \{1, \ldots, m\} \setminus \mathcal{A}$}
                \State {Solve the linear system $Q_{\mathcal{A} \mathcal{A}} \Delta y_{\mathcal{A}} = Q_{\mathcal{A} \mathcal{B}} \phi_{\mathcal{B}}(y^{(\ell)}) - \phi_{\mathcal{A}}(y^{(\ell)})$}
                \State {Compute the remainder of the increment: $\Delta y_{\mathcal{B}} \gets \phi_{\mathcal{B}}(y^{(\ell)})$}
                \State {Combine the increments to form update vector $\Delta y$}
                \State {Update guess: $y^{(\ell + 1)} \gets y^{(\ell)} + \Delta y$}
                \State {Increment $\ell \gets \ell + 1$}
            \Until{convergence}
            \State {\Return $y^{(\ell)}$}
        \EndFunction
    \end{algorithmic}
\end{algorithm}

\section{Variational Quantum and Neural Algorithms for DEM-C Granular Media Simulation}
\label{sec:vqa}

A variational quantum algorithm (VQA) is a hybrid iterative algorithm utilizing both a classical CPU and a noisy intermediate-scale quantum processing unit (QPU), with the two halves operating in tandem. The CPU optimizes a set of parameters $\theta$ that the QPU uses to prepare a parameterized quantum ansatz $\ket{\psi_\theta}$ corresponding to the solution of the desired problem. We use $\ket{\psi_\theta}$ interchangeably to refer to both the state itself and its corresponding state vector. For example, VQAs can be used to identify the ground state of an $n$-qubit Hamiltonian $H$ by optimizing the Rayleigh quotient
\begin{equation}
    L(\theta) = \frac{\langle \psi_\theta | H | \psi_\theta\rangle}{\langle \psi_\theta | \psi_\theta\rangle}
    \label{rayleigh_quotient}
\end{equation}
This type of VQA is referred to as the variational quantum eigensolver (VQE). Note that while quantum state vectors are generally presumed to be normalized, it is useful for \cref{vnls_section} to consider this quotient in full generality. We observe that \cref{rayleigh_quotient} is bounded below by the smallest eigenvalue of $H$, and is minimized if and only if $\ket{\psi_\theta}$ is the associated eigenvector. The VQE can be applied to any problem for which there exists a Hamiltonian $H$ with the solution of the original problem encoded in its ground state.

\subsection{The VQLS Algorithm}
\label{vqls_subsection}
Given an invertible $2^n\times 2^n$ matrix $A$ and a unit vector $\ket{b}$ of corresponding dimension, the VQLS \cite{bravo2019variational} is a VQA designed to solve generalized linear systems of the form
\begin{equation}
A\ket{x}\propto\ket{b} \label{vnls_linear_system}
\end{equation}
by identifying an $n$-qubit ansatz $\ket{\psi_\theta}$ whose state vector is proportional to the solution vector $A^{-1} \ket{b}$.
The VQLS minimizes the Rayleigh quotient \cref{rayleigh_quotient} for the Hamiltonian
\begin{equation}
    H_G = A^\dag \paren{I - \ketbra{b}{b}}A,
    \label{vqls_objective}
\end{equation}
for which the quotient value is a nonnegative measure of the projective overlap between $A\ket{\psi_\theta}$ and $\ket{b}$. In particular, the Rayleigh quotient equals zero if and only if $\ket{\psi_\theta}$ is a solution to \cref{vnls_linear_system}. The trace distance between $\ket{\psi_\theta}$ and $A^{-1}\ket{b}$ is bounded above \cite{bravo2019variational, knitter2022vnls} by $\kappa\sqrt{L(\theta)}/\norm{A}$, where $\kappa$ and $\norm{A}$ are the condition number and operator norm of $A$, and $L(\theta)$ is the Rayleigh quotient loss constructed using \cref{vqls_objective}. For a given linear system, this bound identifies a direct relationship between the loss value and the model accuracy. This objective is known in \cite{bravo2019variational} as the \textit{global} VQLS objective, which we consider as a basis for comparison with the VNLS in section \ref{vnls_section}.

An alternative Rayleigh quotient objective is the \textit{local} VQLS \cite{bravo2019variational}, based on the Hamiltonian \begin{equation}
    H_L = A^\dagger U\paren{I - \frac 1 n \sum_{j=1}^n \ketbra{0_j}{0_j}}U^\dagger A
    \label{vqls_local_objective}
\end{equation} as an alternative to \cref{vqls_objective}. Each summand $\ketbra{0_j}{0_j}$ refers to a local projector operating only on qubit $j$, and $U$ is a unitary operator that maps $\ket{0}$ to $\ket{b}$. The global and local VQLS objectives have identical solutions, though the latter is often more easily trainable, with fewer barren plateaus in its optimization landscape \cite{bravo2019variational}. When utilizing either Hamiltonian, it is common practice to divide the Rayleigh quotient of $H_G$ or $H_L$ by the Rayleigh quotient of $H_A = A^\dagger A$ in constructing the VQLS loss function; this modification improves training for ill-conditioned systems \cite{bravo2019variational}. %We also utilize the hardware-efficient ansatz from \cite{bravo2019variational} as our quantum circuit model.

To execute the VQLS, it is necessary to represent $A$ as a linear combination of unitary operators, each operator corresponding with a separate quantum gate. For our implementation, we rely on the fact that any invertible $2^n\times 2^n$ matrix may be expressed as a complex linear combination of local Pauli strings: \[A = \sum_{i}c_i\paren{A_{i_1}\otimes\cdots\otimes A_{i_n}}\;\text{ for } A_{i_{j}}\in\braces{I,X,Y,Z}.\] In general, the number of terms comprising $A$ is $O(4^n)$; though the VQLS is known to operate well at scale, this requirement does constrain the type of systems it can feasibly solve as the number of qubits increases.

\subsection{The VNLS in Contrast with VQLS}
\label{vnls_section}
We also explore the variational neural linear solver (VNLS) \cite{knitter2022vnls} as a black box solver for the DEM-C simulator. The VNLS is a de-quantization of the VQLS, designed to solve the same generalized linear systems from \cref{vnls_linear_system}. Relying on neural network quantum states (NQS), a quantum-inspired deep learning paradigm for solving many-body problems \cite{carleo2017solving}, the VNLS bypasses the need for a QPU by modeling the state vector of a quantum ansatz $\ket{\psi_\theta}$ using a neural network. In order to make the calculation of \cref{rayleigh_quotient} computationally tractable at scale, NQS reformulates the Rayleigh quotient as a stochastic objective. With a small elaboration from \cite{knitter2022vnls}, we carry out the same process with the Rayleigh quotient of \cref{vqls_objective} to obtain
\begin{equation}
	\label{vnls_objective}
    L(\theta) = \underset{x \sim \pi_\theta}{\mathbb{E}}\bracket{l_\theta(x)},
\end{equation}
where $l_\theta(x)$, referred to as the \textit{local energy}, is defined by
\begin{equation}
    \label{vnls_local_energy}
	l_\theta(x) =
	\frac 1 {\ip{x|\psi_\theta}}\paren{\ip{x|A^\dagger A|\psi_\theta}-\ip{x|A^\dagger|b} \underset{{x'\sim \rho}}{\mathbb{E}}\bracket{\frac{\ip{x'|A|\psi_\theta}}{\ip{x'|b}}}}.
	\end{equation}
Here $x$ and $x'$ refer to state vector indices, encoded in binary as qubit spin configurations, and the distributions $\pi_\theta$ and $\rho$ are defined by
\begin{equation}
    \pi_\theta(x) = \frac{\abs{\ip{x|\psi_\theta}}^2}{\ip{\psi_\theta|\psi_\theta}}\quad
    \text{ and }\quad
    \rho(x')=\abs{\ip{x'|b}}^2.
\end{equation}
Thus we may estimate $L(\theta)$ by calculating local energy values with respect to index batches sampled from $\pi_\theta$ and $\rho$ using Monte Carlo methods. The original NQS architecture utilizes a restricted Boltzmann machine to model $\ket{\psi_\theta}$ and trains using an analogue of natural gradient descent known as stochastic reconfiguration (SR) \cite{carleo2017solving, sorella_aps98}. It is this version that we utilize within the DEM-C simulator.

The VNLS can operate on matrices stored using the local Pauli decomposition from \cref{vqls_subsection}, but since this fully classical architecture does not have the same input format requirements inherent to the VQE, the implementation of VNLS we use stores the entries of $A$ directly in a compressed sparse row (CSR) format, which is more efficient for the problem sizes we consider.

\begin{figure}
    \centering
\includegraphics[width=\textwidth]{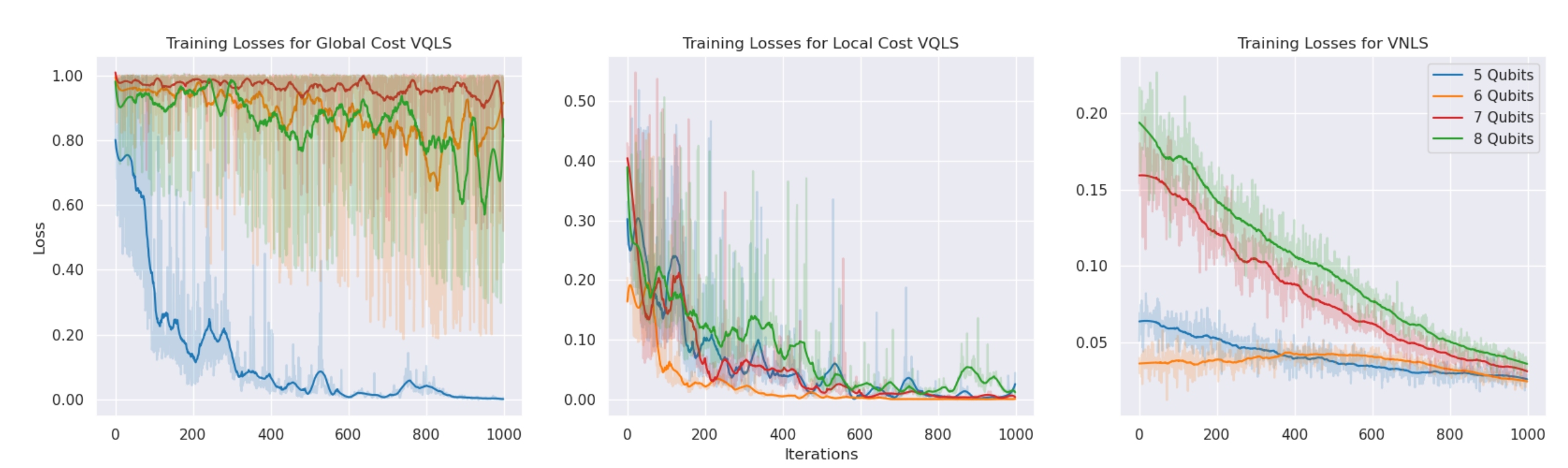}
  \caption{\em Comparisons between the global cost VQLS (left), local cost VQLS (center) and the VNLS (right) on the $\kappa=10$ Ising-inspired problem. Loss curves are de-noised using a Savitzky--Golay filter; the true loss curves are also presented here, with lower opacity.}
    \label{fig:vqls_vnls_baseline_comp}
\end{figure}

\section{Numerical Results}
\label{sec:result}
We first give a brief overview of baseline testing for both the VQLS and VNLS paradigms on well-conditioned, Ising-inspired linear systems. Afterward, we highlight some exploratory results that suggest the applicability of these solvers to linear systems derived from DEM-C simulations, particularly by assessing the performance of the VNLS on solving complementarity problems arising from simulation of 100 colliding spheres. Out of considerations of practicality, a few concessions were made during testing. Firstly, at this stage, we only assess the performance of the VNLS and not the VQLS, since the general method for encoding Hamiltonians into quantum circuits---Pauli string decompositions---does produce a considerable, though subexponential, computational cost that is not experienced by the fully classical VNLS. Secondly, at this time we do not incorporate frictional forces into the DEM-C simulation, instead modeling frictionless collisions. %Elaborate on limitation of systems due to Pauli decomposition for VQLS. Include lack of friction in testing.

\subsection{Baseline Testing of the VQLS and the VNLS}
\label{sec:baseline_results}
As introduced in \cite{bravo2019variational}, we rely on the Ising-Inspired VQLS problem, a highly row-sparse system with a user-specified condition number $\kappa$, for baseline analysis of the VQLS. Fig. \ref{fig:vqls_vnls_baseline_comp} illustrates the performance of the VQLS on $\kappa=10$ Ising-inspired problems for 5,6,7, and 8 qubit systems, which were simulated using Qiskit's Aer backend simulator. Optimization was performed in all cases by quantum natural SPSA, a gradient-free analogue of quantum natural gradient descent. Each quantum circuit was sampled for 1000 shots in order to construct the appropriate expectation values. Solver performance using the global cost function did appear to drop off with problem size, highlighting the barren plateau issues discussed in \cite{bravo2019variational}. The local cost function did not present the same issues, adequately learning the true solution for all sizes shown.

Analogously, and as done in \cite{knitter2022vnls}, we also perform baseline testing of the VNLS using the Ising-inspired problem, whose results are also given in \cref{fig:vqls_vnls_baseline_comp}. Here, we present the performance of the VNLS, trained using SR, on the same systems used to test the VQLS. Though the VNLS cost function is analogous to the global VQLS cost, we do note that the architectures are sufficiently different that one cannot make a one-to-one, iteration-by-iteration comparison between the two loss curves. Nonetheless, we may observe that the VNLS is able to identify the correct solution state with sufficient accuracy, performing better than the global VQLS. %One particular advantage of the VNLS over the VQLS is that the variance of the stochastic loss approaches zero as the loss itself approaches zero, allowing for more accurate optimization further along the learning process \cite{knitter2022vnls}.

\subsection{DEM-C Testing Using the VNLS}

\begin{figure}
    \centering \includegraphics[width=0.85\textwidth]{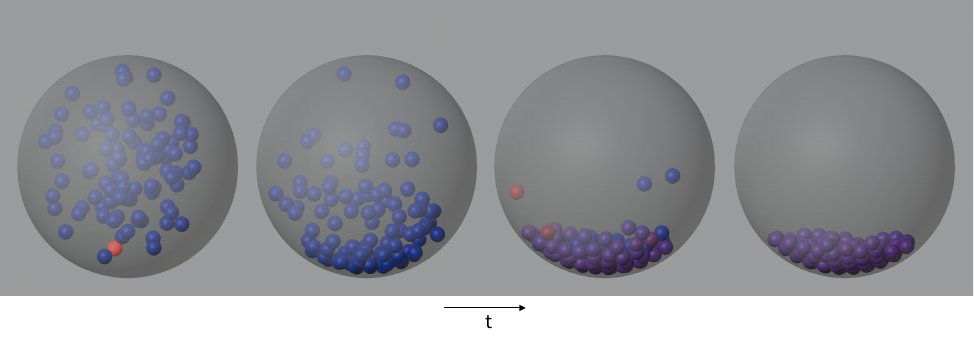}
    \caption{\em Snapshots from a simulation of one hundred spheres sedimenting under gravity inside a spherical enclosure. The inter-particle collisions are resolved using the LCP formulation presented in Section \ref{sec:lcp}. Color represents the magnitude of the contact force, with red indicating high and blue indicating zero.}
    \label{fig:demo_lcp}
\end{figure}

\begin{figure}
    \centering
    \begin{subfigure}[b]{0.6\textwidth}
        \includegraphics[width=\textwidth]{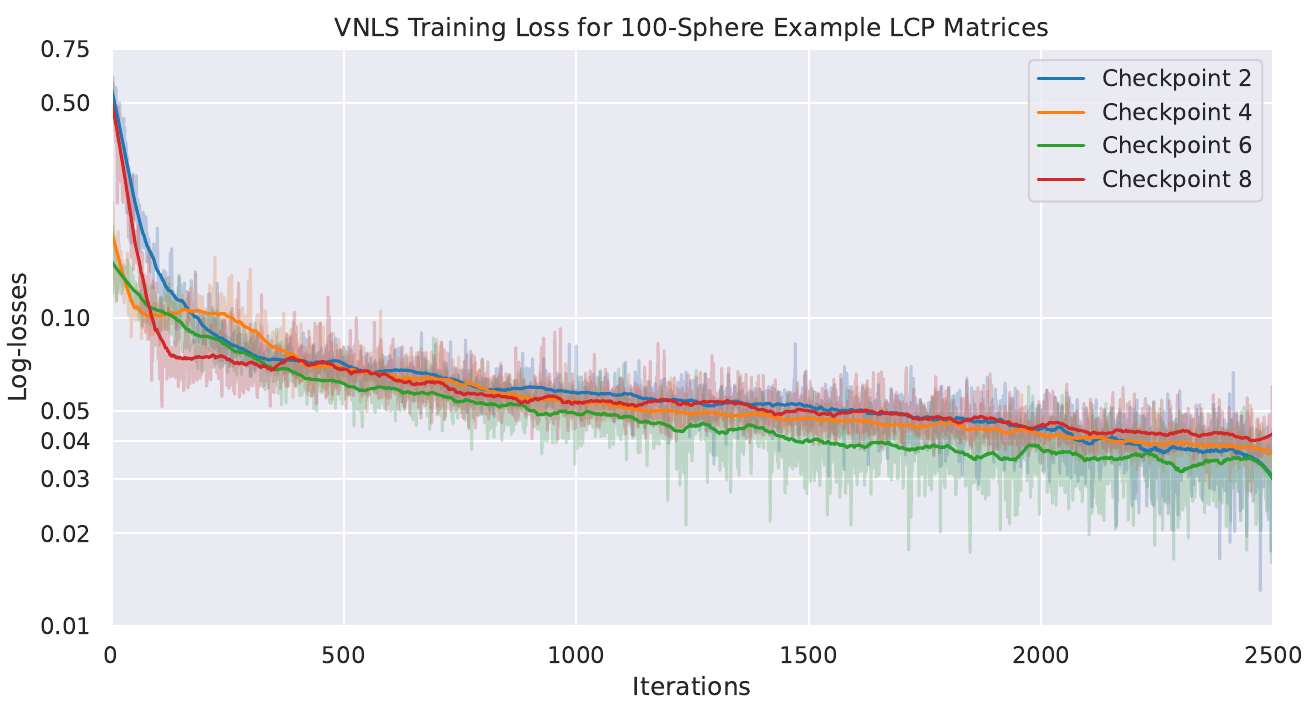}
        \caption{\em Log-loss curves over 2500 iterations, de-noised with a Savitzky--Golay filter.}
        \label{fig:vnls_lcp}
    \end{subfigure}
    \qquad
    \begin{subfigure}[b]{0.3\textwidth}
        \centering
        \begin{tabular}{ccc}
            \hline %$\quad$ \\
            Checkpoint & $\quad n\quad$ & $\kappa$  \\ %$\quad$ \\
            \hline %$\quad$ \\
            % 2 & 8 & $20.3975$ \\ $\quad$ \\
            % 4 & 8 & $40.5105$ \\ $\quad$ \\
            % 6 & 8 & $74.5506$ \\ $\quad$ \\
            % 8 & 9 & $329.6981$ \\ $\quad$ \\
            % \hline $\quad$ \\
             2 & 8 & $20.4$ \\ %$\quad$ \\
            4 & 4 & $40.5$ \\ %$\quad$ \\
            6 & 8 & $74.6$ \\% $\quad$ \\
            8 & 9 & $329.7$ \\ %$\quad$ \\
            \hline \vspace{0.3in}
    
        \end{tabular} 
        \caption{\em The qubit and condition numbers for each curve in \cref{fig:vnls_lcp}.}
        \label{tab:example_kappas}
    \end{subfigure}
    \caption{\em Demonstration of a complex RBM VNLS ansatz solving linear systems derived from the DEM-C LCP at several moments in time of a 100-sphere system. Matrices were generated at intermediate steps of the minimum map Newton solver, at points in time separated by 40,000 Euler time step updates. The dimension of each system must be padded to the nearest power of two to make it amenable to VNLS/VQLS.}
\end{figure}

%\begin{table}[ht!]
%    \centering
%    \begin{tabular}{ccccc}
%        \toprule
%         Checkpoint & 2 & 4 & 6 & 8  \\
%         \midrule
%        $\kappa$ & $20.3975$ & $40.5105$ & $74.5506$ & $329.6981$ \\
%         \bottomrule
%    \end{tabular}
%    \caption{Condition numbers of the linear systems depicted in \cref{fig:vnls_lcp}}   
%\end{table}

Our experiments were performed using a rudimentary granular medium simulation that models a number of small rigid spheres contained within an impermeable spherical container. These simulations do not reflect the full generality of the standard DEM-C framework: due to limitations in computational capacity, we ignore frictional forces and focus on these solvers' general ability to serve within the time evolution framework. Likewise, we have only performed simulations using the VNLS.

We model a system of 100 rigid spheres, free falling under the influence of gravity inside a hollow spherical shell, using a Newton-based complementarity solver as discussed in \cref{sec:lcp}.

A visual depiction of such a system is given by figure \ref{fig:demo_lcp}. Figure \ref{fig:vnls_lcp} gives a demonstration of VNLS, using a complex RBM as an ansatz, as it solves a series of linear systems produced at different time steps of a 100-sphere simulation. These linear systems were specifically chosen at time steps separated by 40,000 Euler updates; more specifically, each time step requires multiple iterations in the minimum map Newton solver to identify the relevant velocities, and the systems represent intermediate iterations of this minimum map solver at each point in time.

It is necessary to pad the dimension of the linear system to the next largest power of 2 for it to fit the form of a quantum system, which may be done easily without altering the solution of the problem by setting the padded diagonal entries of $A$ to 1, with all remaining padded entries of $A$ and $\ket{b}$ set to 0. For ease of comparison, we also scaled the matrices of these systems to have unit 2-norm, bringing them closer in line with the Ising-inspired baselines from \cref{sec:baseline_results}. We do note in \cref{tab:example_kappas} that the condition numbers of these matrices increase over time, which is known to influence the accuracy guarantees of both the VQLS and VNLS. For these examples, we trained the RBM ansatze for 2500 iterations each, with 1024 Monte Carlo samples taken per iteration, using stochastic reconfiguration \cite{sorella_aps98} with a learning rate of $0.05$. The estimated loss values are noisy due to the nature of the optimization, but the de-noised trends indicate that the VNLS is successfully optimizing the stochastic loss function in all cases.

\subsection{Challenges in Using VQLS for DEM-C}

The VNLS is a direct de-quantization of the VQLS, and as discussed in \cref{sec:baseline_results}, they perform comparably well on Ising-inspired baseline problems.
However, there are practical barriers to naively using VQLS to solve linear systems arising from DEM-C simulations.
This class of linear systems, being generally sparse with efficiently retrievable entries, satisfy the required conditions \cite{knitter2022vnls, nest2009simulating} for effective use in the quantum-inspired VNLS.
In contrast, the VQLS---a truly quantum algorithm---requires that these matrices be encoded into a form amenable to quantum measurement, such as Pauli strings.
The linear systems from frictionless LCP formulation are symmetric and comprise of real numbers, hence their Pauli decompositions may consist of up to $\frac{1}{2} (4^n + 2^n)$ terms in $n$-qubit systems.
Direct computation of these expansion coefficients via the fast Walsh-Hadamard transform \cite{georges2025pauli}, without leveraging the sparsity structure, requires $O(n 4^n)$ elementary arithmetic operations.
Thus, the Pauli string decomposition of DEM-C matrices may itself be a computational bottleneck for effective use of the VQLS algorithm.

\begin{figure}
    \centering
    \begin{subfigure}{0.32\textwidth}
        \centering
        \includegraphics[width=\linewidth]{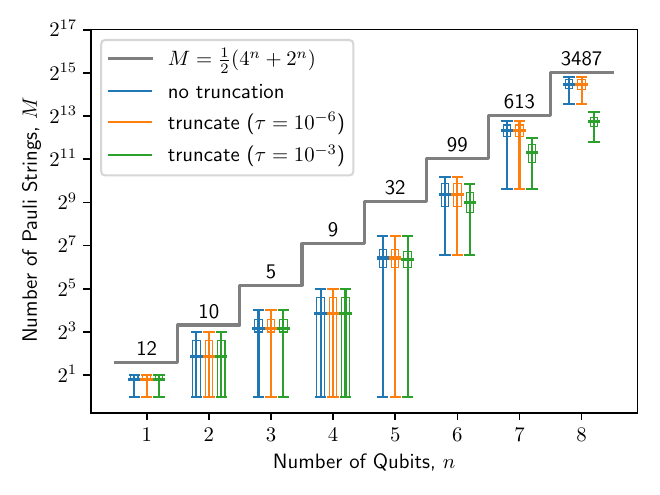}
        \caption{\textit{64-sphere system}}
    \end{subfigure}
    ~
    \begin{subfigure}{0.32\textwidth}
        \centering
        \includegraphics[width=\linewidth]{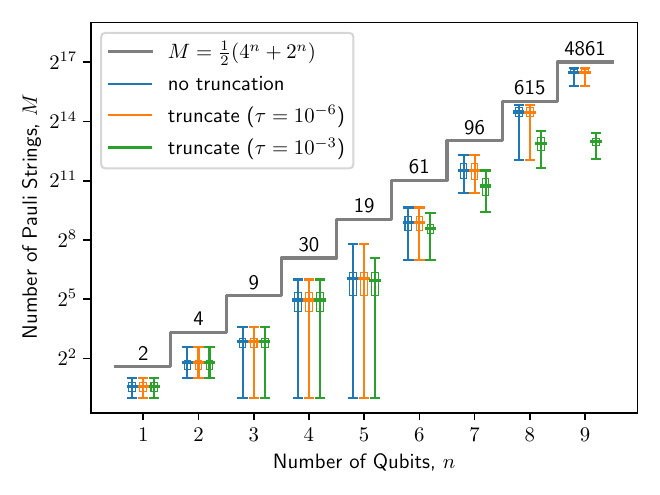}
        \caption{\textit{125-sphere system}}
    \end{subfigure}
    ~
    \begin{subfigure}{0.32\textwidth}
        \centering
        \includegraphics[width=\linewidth]{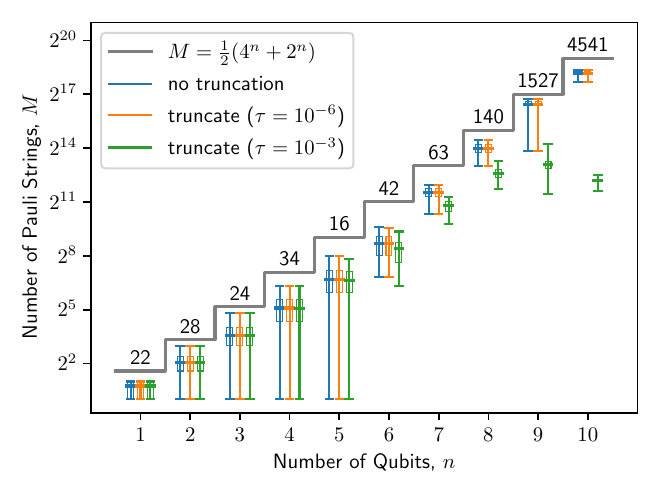}
        \caption{\textit{216-sphere system}}
    \end{subfigure}
    \caption{\textit{%
        Pauli-string decomposition of frictionless LCP linear systems over the course of three different DEM-C simulations.
        We plot the number of terms in the Pauli expansions as a function of number of qubits needed to represent the matrices in blue.
        We also overlay the number of terms when the Pauli expansion is truncated by discarding coefficients with absolute values smaller than $10^{-6}$ (red) and $10^{-3}$ (green).
        The gray stair-plot corresponds to the theoretically maximum number of terms in the Pauli expansion.
        The numbers above this stair-plot indicate how many of the linear systems over the course of the simulation were expressed as $n$-qubit systems and the box-plot levels represent the corresponding minimum, maximum, 25\% and 75\% quantiles, and average number of Pauli terms.
    }}
    \label{fig:pauli}
\end{figure}

Figure~\ref{fig:pauli} illustrates a preliminary analysis of the Pauli decomposition step for DEM-C linear systems.
Over the course of three rigid body simulations consisting of 64, 125, and 216 spheres, we use the open-source \texttt{Qiskit} library to compute the number of terms in the Pauli strings decomposition of the normalized and padded $2^n \times 2^n$ linear systems, optionally dropping the terms whose coefficients are smaller than a pre-specified tolerance $\tau$ in absolute values.
Without any truncation, the number of Pauli terms closely follow the theoretical maximum, but we do observe a sharp decrease in these numbers when the tolerance is loose ($\tau = 10^{-3}$), especially for the larger simulations and higher number of qubits.
These observations suggest that VQLS may be feasible if we are willing to accept some approximation error to the DEM-C linear systems.
However, how this allowance affects the overall accuracy of rigid body simulations, especially in the context of noisy quantum circuit operations, requires further research.
In addition, significant recent work has been done to reduce measurement overhead in the case of energy estimation through exploitation of simultaneously measurable Pauli strings \cite{gokhale2020optimization}, or through classical shadows \cite{huang2020predicting}. The adaptation of similar methods to the VQLS framework is a worthwhile, but nontrivial endeavor, exceeding the scope of this work.

\section{Conclusions and Future Work}
\label{sec:conclusion}

In this paper, we give a general description of a practically-motivated area of application, DEM-C granular medium simulation, in which both quantum and quantum-inspired machine learning methods may alleviate computational costs that prevent standard methods from achieving efficient scaling. As a specific example, we explore the utility of the variational quantum linear solver and its de-quantization, the variational neural linear solver, as black box linear system solvers within a Newton-based method for solving linear complementarity problems, a vital component of the DEM-C simulation that also serves as the primary computational bottleneck. Our experiments find that the VNLS is amenable to incorporation inside such a framework, as demonstrated on examples generated from simplified systems of up to 100 colliding bodies. On the other hand, the VQLS algorithm suffers from practical considerations of the number of Pauli strings needed to represent the system---a limitation the VNLS can more easily bypass. Preliminary results show that the growth in the number of Pauli strings could be constrained if we truncate the expansion to moderate tolerances; this reduction seem to be more prominent for larger simulations. However the ramifications of approximating the linear systems remains unclear, especially in the context of near-term noisy quantum hardware. Nonetheless, this work provides an insight into potential use cases, within the realm of physical simulation, for quantum and quantum-inspired machine learning algorithms.

One potential direction for future work, in addition to expanding the simulation to model larger numbers of bodies under more complicated forces, would be to introduce a non-granular component for imparting external forces to the colliding granules. This component would serve to better depict the type of practical applications that these simulations would be used to model---in the case of ground vehicle mobility modeling, such a component might be the wheel of a vehicle as it moves over terrain. At the same time, there exist a few interesting areas of further exploration on the architectural side. An interesting line of research is currently underway exploring non-stochastic optimization strategies for RBM NQS ansatze \cite{li2024improved}. These methods bypass the need for random sampling, and could serve as an interesting potential avenue for further improving the practical scalability of the VNLS as an LCP solver.

Moving away from RBMs, significant progress has been made, particularly within the domain of \textit{ab initio} quantum chemistry for developing scalable NQS ansatze based on autoregressive neural networks \cite{sharir2020deep, wu2023nnqstransformer, zhao2021overcoming}. Unlike RBMs, which can only learn unnormalized quantum states and require approximate Monte Carlo sampling methods that do not parallelize well, these networks are constrained to model normalized quantum states. As a result, they allow for exact, parallelizable sampling from the state vector distribution. Furthermore, they do not require SR for effective training, and typically perform well with first-order optimizers that approximate second-order information, like Adam. For these reasons, we encourage the exploration of autoregressive ansatze in domains of application beyond computational quantum chemistry, and our VNLS-based DEM-C simulator outlines an avenue to explore their capabilities for physical simulation.

\section{Acknowledgements}
We acknowledge support from
the Automotive Research Center at the University of Michigan (UM) in accordance with Cooperative Agreement W56HZV-19-2-0001 with U.S. Army DEVCOM Ground Vehicle Systems Center.

Sandia National Laboratories is a multi-mission laboratory managed and operated by National Technology \& Engineering Solutions of Sandia, LLC (NTESS), a wholly owned subsidiary of Honeywell International Inc., for the U.S. Department of Energy’s National Nuclear Security Administration (DOE/NNSA) under contract DE-NA0003525. This written work is authored by an employee of NTESS. The employee, not NTESS, owns the right, title and interest in and to the written work and is solely responsible for its contents. This paper describes objective technical results and analysis. Any subjective views or opinions that might be expressed in the paper do not necessarily represent the views of the U.S. Department of Energy or the United States Government. The publisher, by accepting the article for publication, acknowledges that the United States Government retains a non-exclusive, paid-up, irrevocable, world-wide license to publish or reproduce the published form of this article or allow others to do so, for United States Government purposes. The DOE will provide public access to these results of federally sponsored research in accordance with the DOE Public Access Plan \texttt{https://www.energy.gov/downloads/doe-public-access-plan}.

\appendix

\section{DEM-C Formulation of Rigid Body Contact}

\subsection{Derivation of Collision Force in Global Frame}
\label{app:fcol_global}

The evolution of $p_i, v_i \in \mathbb{R}^3$, which denote the position and velocity of the center of mass of the $i$-th sphere, is given by Newton's second law of motion:
\begin{equation*}
    \frac{\D p_i}{\D t} = v_i, \quad m_i \frac{\D v_i}{\D t} = f_i + f_i^\text{col},
\end{equation*}
where $f_i$ and $f_i^\text{col}$ are the external and collision force acting on the $i$-th sphere. Consider the $k$-th contact between the $i_k$- and $j_k$-th spheres, and let $\nu_k, \tau_{k,1}, \tau_{k,2}$ be the orthonormal coordinate frame local to the contact: $\nu_k$ is normal to the contact plane and points toward the $j_k$-th sphere while $\tau_{k,1}, \tau_{k,2}$ span the contact plane. We decompose the contact force acting on the $j_k$-th sphere due to the $k$-th contact in this local coordinate frame as \begin{equation*}
    f_{(k)}^\text{col} = \gamma_{k,n} \nu_k + \gamma_{k,1} \tau_{k,1} + \gamma_{k,2} \tau_{k,2};
\end{equation*}
an equal and opposing force $-f_{(k)}^\text{col}$ is active on the $i_k$-th sphere by Newton's third law. We obtain the contact force on the $i$-th sphere by summing the contribution from all collision pairs:
\begin{equation}
    \label{eq:fcol_local}
    f_i^\text{col}
    = \sum_{\substack{k = 1\\j_k = i}}^{n_c} f_{(k)}^\text{col} + \sum_{\substack{k = 1\\i_k = i}}^{n_c} (- f_{(k)}^\text{col})
    = \sum_{\substack{k = 1\\j_k = i}}^{N_c} (\gamma_{k,\nu} \nu_k + \gamma_{k,1} \tau_{k,1} + \gamma_{k,2} \tau_{k,2}) - \sum_{\substack{k = 1\\i_k = i}}^{n_c} (\gamma_{k,\nu} \nu_k + \gamma_{k,1} \tau_{k,1} + \gamma_{k,2} \tau_{k,2}).
\end{equation}
To derive the collision force in the global coordinates, we define the $\text{L2G}_k$ operator that maps a vector $y \in \mathbb{R}^3$ in the local coordinate frame of the $k$-th contact to a global vector $z = \text{L2G}_k(y) \in \mathbb{R}^{3 n_b}$. The components of the output vector $z = (z_1, \ldots, z_{3 n_b})$ of this operation is related to the components of the input vector $y = (y_1, y_2, y_3)$ as
\begin{equation*}
    z_\ell =
    \begin{cases}
        -y_{\ell + 3 - 3 i_k} & \text{if} \quad 3 i_k - 2 \leq \ell \leq 3 i_k \\
        y_{\ell + 3 - 3 j_k} & \text{if} \quad 3 j_k - 2 \leq \ell \leq 3 j_k \\
        0 & \text{otherwise}.
    \end{cases}
\end{equation*}
Let $d_{k,\nu} = \text{L2G}_k(\nu_k)$, $d_{k,1} = \text{L2G}_k(\tau_{k,1})$, and $d_{k,2} = \text{L2G}_k(\tau_{k,2})$ be the global direction vectors associated with the local frame. Then, given the expression of global collision force in \cref{eq:fcol_global},
\begin{equation*}
    f^\text{col} = \sum_{k = 1}^{n_c} (\gamma_{k,\nu} d_{k,\nu} + \gamma_{k,1} d_{k,1} + \gamma_{k,2} d_{k, 2}),
\end{equation*}
it is straightforward to verify that $f^\text{col}(3 i - 2 : 3 i)$, the segment of the global collision force vector corresponding to sphere $i$, agrees with \cref{eq:fcol_local}.

We use this $\text{L2G}_k$ operator to also define the generalized vectors corresponding to the linearized direction vectors of the $k$-th contact: $\widehat{d}_{k,\ell} = \text{L2G}_k(\widehat{\tau}_{k,\ell})$ for $1 \leq \ell \leq s$.

\subsection{Dissipation Formulation of Coulomb Friction}
\label{app:coulomb}

The Coulomb model imposes two restrictions on the frictional contact force: (i) the maximum magnitude of the frictional force is proportional to the magnitude of the normal contact force, and (ii) the frictional force opposes the relative motion between the pair of particles. Given coefficient of friction $\mu$, the first condition yields an inequality constraint,
\begin{equation*}
    \sqrt{\gamma_{k,1}^2 + \gamma_{k,2}^2} \leq \mu \gamma_{k,\nu},
\end{equation*}
that defines the Coulomb friction cone. To obtain a mathematical expression for the second condition, we observe that for a given vector $y \in \mathbb{R}^3$ and the generalized velocity vector $v = (v_1, \ldots, v_{n_b})$ of the system, we have
\begin{equation*}
    \text{L2G}_k(y)^\top v = -y^\top v_{i_k} + y^\top v_{j_k} = y^\top (v_{j_k} - v_{i_k}).
\end{equation*}
It follows that
\begin{equation*}
    \begin{split}
        (\gamma_{k,1} d_{k,1} + \gamma_{k,2} d_{k,2})^\top v
        &= \gamma_{k,1} d_{k,1}^\top v + \gamma_{k,2} d_{k,2}^\top v \\
        &= \gamma_{k,1} \text{L2G}_k(\tau_{k,1})^\top v + \gamma_{k,2} \text{L2G}_k(\tau_{k,2})^\top v \\
        &= \gamma_{k,1} \tau_{k,1}^\top (v_{j_k} - v_{i_k}) + \gamma_{k,2} \tau_{k,2}^\top (v_{j_k} - v_{i_k}) \\
        &= (\gamma_{k,1} \tau_{k,1} + \gamma_{k,2} \tau_{k,2})^\top (v_{j_k} - v_{i_k}) \\
        &= f_{(k)}^{\text{fric},\top} (v_{j_k} - v_{i_k}).
    \end{split}
\end{equation*}
For a given velocity $v$ of the system, minimizing this vector dot product ensures that the friction force $f_{(k)}^\text{fric}$ points in the opposite direction to the relative velocity $v_{j_k} - v_{i_k}$ between the colliding particles projected onto the contact plane. We have reformulated the Coulomb friction model as a maximal dissipation law stated in \cref{eq:coulomb_dissipation}:
\begin{equation*}
    (\gamma_{k,1}, \gamma_{k,2}) = \operatorname*{argmin}_{\sqrt{\widehat{\gamma}_{k,1}^2 + \widehat{\gamma}_{k,2}^2} \leq \mu \gamma_{k,\nu}} (\widehat{\gamma}_{k,1} d_{k,1} + \widehat{\gamma}_{k,2} d_{k,2})^\top v.
\end{equation*}

This property of the $\text{L2G}_k$ operator is also used in \cref{sec:lcp_formulation} to derive a complementarity condition. Note that
\begin{equation*}
    d_{k,\nu}^\top v = \text{L2G}_k(\nu_k)^\top v = \nu_k^\top (v_{j_k} - v_{i_k}),
\end{equation*}
i.e., $d_{k,\nu}^\top v$ is the magnitude of the relative velocity between the particles of the $k$-th contact along the normal direction. It is zero when the particles are colliding, and positive after the contact ends.

\bibliographystyle{plain}
\bibliography{references}
\end{document}